\documentclass[preprint,12pt]{elsarticle}

\usepackage{graphicx}

\usepackage{amssymb}
\usepackage{bm}

\journal{Physics Letters B}

\begin{document}

\begin{frontmatter}



\title{Magnetic neutrino scattering on atomic electrons revisited}


\author[KAK]{Konstantin A. Kouzakov\corref{cor1}}
\ead{kouzakov@srd.sinp.msu.ru}
\author[AIS1,AIS2]{Alexander I. Studenikin}
\ead{studenik@srd.sinp.msu.ru}

\address[KAK]{Department of Nuclear Physics and Quantum Theory of Collisions, Faculty of Physics, Moscow State University, Moscow 119991, Russia}
\address[AIS1]{Department of Theoretical Physics, Faculty of Physics, Moscow State University, Moscow 119991, Russia}
\address[AIS2]{Joint Institute for Nuclear Research, Dubna 141980, Moscow Region, Russia}
\cortext[cor1]{Corresponding author}

\begin{abstract}
We reexamine the role of electron binding effects in the inelastic
neutrino-atom scattering induced by the neutrino magnetic moment.
The differential cross section of the process is presented as a
sum of the longitudinal and transverse components, according to
whether the force that the neutrino magnetic moment exerts on
electrons is parallel or perpendicular to momentum transfer. The
atomic electrons are treated nonrelativistically. On this basis,
the recent theoretical predictions concerning the magnetic
neutrino-impact ionization of atoms are critically discussed.
Numerical calculations are performed for ionization of a
hydrogenlike Ge$^{+31}$ ion by neutrino impact.
\end{abstract}
\end{frontmatter}
%


%
\section{Introduction}
Neutrino electromagnetic properties are among the most intriguing
issues in neutrino physics (see Ref.~\cite{giunti09} for a review
on this subject). Within the Standard Model the value of the
neutrino magnetic moment in units of the Bohr magneton
$\mu_B=e/(2m_e)$ is~\cite{marciano77,lee77,fujikawa80}
$$
\mu_\nu\approx3.2\times10^{-19}\left(\frac{m_\nu}{1~{\rm
eV}}\right),
$$
where $m_e$ and $m_\nu$ are the electron and neutrino masses,
respectively. Therefore, experimental evidence for a value which
is much larger than $10^{-19}$ will signalize new physics.

The nowaday laboratory searches for neutrino magnetic moments are
dealing with reactor (anti)neutrino sources (MUNU~\cite{munu05},
TEXONO~\cite{texono07}, and GEMMA~\cite{gemma10}). The best upper
limit for $\mu_\nu$ obtained so far in the reactor experiments is
$\mu_\nu\leq3.2\times10^{-11}$~\cite{gemma10} (see also references
in the review article~\cite{giunti09}), which still exceeds by an
order of magnitude the most stringent astrophysical constraint
$\mu_\nu\leq3\times10^{-12}$~\cite{raffelt90}.
Since the differential cross section (DCS) for the magnetic
scattering of a neutrino with energy $E_\nu$ on a free electron
(FE) at energy transfer $T$ is given
by~\cite{domogatskii71,vogel89}
\begin{equation}
\label{dcs_FE}\frac{d\sigma_{(\mu)}}{dT}=\frac{\pi\alpha^2\mu_\nu^2}{m_e^2}\left(\frac{1}{T}-\frac{1}{E_\nu}\right),
\end{equation}
where $\alpha$ is the fine-structure constant, a way to enhance
the sensitivities of the experiments is to reduce the low-energy
threshold of the detectors with respect to the deposited energy
$T$. Note in this connection that the DCS due to the weak
interaction, $d\sigma_{(w)}/dT$, is practically constant in $T$
when $E_\nu\gg T$~\cite{vogel89}.

The electrons of the detector material are bound to atoms and
therefore the FE picture is applicable only when
$T\gg\varepsilon_b$, where $\varepsilon_b$ is the electron binding
energy. If $T\sim\varepsilon_b$, the electron binding effects must
be taken into consideration. Recently, it was claimed that at
$T\sim\varepsilon_b$ the DCS of the inelastic magnetic neutrino
scattering on atomic electrons (AE) strongly increases, as
compared to the FE case~(\ref{dcs_FE}), owing to an atomic
ionization effect~\cite{wong10}. This means that the sensitivity
of the reactor experiments searching for $\mu_\nu$ can be
significantly enhanced. As demonstrated below, the authors of
Ref.~\cite{wong10} incorrectly evaluated the DCS, groundlessly
employing the approximation of real photons in the whole range of
the final neutrino angles. In this respect, in a more recent
work~\cite{voloshin10} it was argued, by means of a
quantum-mechanical sum rule, that when $E_\nu\gg T$ and $T$ is
comparable to the characteristic atomic energies the values of DCS
for the AE and FE cases are almost the same. While this promising
analytical method yields a physically reasonable conclusion for
the one-electron case, we show in the analysis carried out below
that it requires further elaboration in order to be correctly
applied to many-electron atomic systems.

In the present work, we formulate a new theoretical approach which
allows us to consider the magnetic neutrino scattering on atomic
electrons in a consistent and physically transparent fashion. This
approach can be traced down to that developed for the penetration
of relativistic charged projectiles in matter~\cite{fano63}.
According to it, the DCS is given by the sum of two physically
distinct components--the longitudinal and the transverse, which
are associated with the corresponding components of the force
imposed by the neutrino on electrons with respect to the direction
of the momentum transfer ${\bf q}$. Such a decomposition also
enables us to clearly distinguish between the contributions from
excited atomic states taken into account in Refs.~\cite{wong10}
and~\cite{voloshin10}, respectively, and to inspect consistently
the results of those studies. The developed theory is then
numerically realized in the case of magnetic \emph{neutrino-impact
ionization} ($\nu II$) of a hydrogen-like Ge$^{+31}$ ion. This
specific case models the $K$-shell ionization of Ge by neutrino
impact, which is relevant to the reactor
experiments~\cite{texono07,gemma10}.

The article is organized as follows. Sec.~\ref{theory} delivers
general theory for the magnetic neutrino scattering on atomic
electrons as well as a critical account of the theoretical studies
published recently~\cite{wong10,voloshin10}. Numerical
calculations are presented and analyzed in Sec.~\ref{res}. The
conclusions are drawn in Sec.~\ref{concl}. The units $\hbar=c=1$
are used throughout unless otherwise stated.

\section{Theory of magnetic neutrino-impact excitation and ionization of atoms}
\label{theory}
We consider the process where a neutrino with energy $E_\nu$ and
momentum ${\bf p}_\nu$ scatters on an atom at energy-momentum
transfer $q=(T,{\bf q})$. In what follows we focus on the role
that plays in this process the interaction of the neutrino
magnetic moment with atomic electrons, assuming the nucleus to be
infinitely massive. The latter assumption is reasonable if $T\gg
E^2_\nu/M$, where $M$ is the nuclear mass, in which case the
recoil of the atom as a whole can be safely neglected and the
coherent neutrino-atom collisions (without changing the atomic
state) are of no relevance. The atomic target is supposed to be
unpolarized and in its ground state $|0\rangle$ with the
corresponding energy $E_0$. We also suppose that $T\ll m_e$ and
$\alpha Z\ll1$, where $Z$ is the nuclear charge, so that both the
initial and the final electronic systems can be treated
nonrelativistically. The incident and final neutrino states are
described by the Dirac spinors assuming $m_\nu\approx0$.

\subsection{The differential cross section}
\label{dcs}
In the low-energy limit the neutrino electromagnetic vertex
associated with the neutrino magnetic moment has the form
\begin{equation}
\label{el-m_vertex}\Lambda_{(\mu)}^i=\frac{\mu_\nu}{2m_e}\sigma^{ik}q_k.
\end{equation}
It should be noted that the $\mu_\nu$ related contribution to the
neutrino-atom scattering does not interfere with that due to weak
interaction because in contrast to the latter it couples neutrino
states with different helicities. Employing first-order
perturbation theory and using the photon propagator in the Coulomb
gauge, the transition matrix element for the considered process
according to Eq.~(\ref{el-m_vertex}) is given by
\begin{eqnarray}
\label{matr_el}M_{fi}^{(\mu)}&=&\frac{2\pi\alpha\mu_\nu}{m_e{\bf
q}^2}(\bar{u}_{{\bf p}_\nu-{\bf q}\lambda_f}u_{{\bf
p}_\nu\lambda_i})\left\{(2E_\nu-T)\langle n|\sum_{j=1}^Ze^{i{\bf
q}\cdot{\bf
r}_j}|0\rangle\right.\nonumber\\
&{}&\left.+\frac{2{\bf q}\times[{\bf p}_\nu\times{\bf
q}]}{T^2-{\bf q}^2}\cdot\langle
n|\sum_{j=1}^Z{\bm\alpha}_je^{i{\bf q}\cdot{\bf
r}_j}|0\rangle\right\},
\end{eqnarray}
where $u_{{\bf p}\lambda}$ is the spinor amplitude of the neutrino
state with momentum ${\bf p}$ and helicity $\lambda$,
${\bm\alpha}_j$ the current operator of the $j$th electron, and
$|n\rangle$ the final atomic state. Using Eq.~(\ref{matr_el}), the
double differential cross section can be presented as
\begin{equation}
\label{DCS}\frac{d^2\sigma_{(\mu)}}{dTdQ}=\left(\frac{d^2\sigma_{(\mu)}}{dTdQ}\right)_\parallel+\left(\frac{d^2\sigma_{(\mu)}}{dTdQ}\right)_\perp,
\end{equation}
where  $Q={\bf q}^2$, $T^2\leq{Q}\leq(2E_\nu-T)^2$, and
\begin{eqnarray}
\label{DCS_paral}\left(\frac{d^2\sigma_{(\mu)}}{dTdQ}\right)_\parallel&=&\frac{\pi\alpha^2\mu_\nu^2}{m_e^2}
\frac{(2E_\nu-T)^2}{4E_\nu^2Q}\left(1-\frac{T^2}{Q}\right)\nonumber\\&{}&\times
\sum_n|\langle n|\rho(-{\bf q})|0\rangle|^2\delta(T-E_n+E_0),
\end{eqnarray}
\begin{eqnarray}
\label{DCS_perp}\left(\frac{d^2\sigma_{(\mu)}}{dTdQ}\right)_\perp&=&\frac{\pi\alpha^2\mu_\nu^2}{m_e^2}
\frac{(2E_\nu-T)^2}{4E_\nu^2Q}
\left[1-\frac{Q}{(2E_\nu-T)^2}\right]\nonumber\\&{}&\times
\sum_n|\langle
n|\hat{{\bf e}}_\perp\cdot{\bf j}(-{\bf q})|0\rangle|^2\delta(T-E_n+E_0).
\end{eqnarray}
In Eqs.~(\ref{DCS_paral}) and~(\ref{DCS_perp}), $\rho(-{\bf q})$
and ${\bf j}(-{\bf q})$ are the Fourier transforms of the electron
density and current density operators, respectively,
\begin{equation}
\label{density}\rho(-{\bf q})=\sum_{j=1}^Ze^{i{\bf q}\cdot{\bf
r}_j},
\end{equation}
\begin{equation}
\label{current}{\bf j}(-{\bf
q})=-\frac{i}{2m_e}\sum_{j=1}^Z\left(e^{i{\bf q}\cdot{\bf
r}_j}\nabla_j+\nabla_je^{i{\bf q}\cdot{\bf r}_j}\right),
\end{equation}
and the unit vector $\hat{{\bf e}}_\perp$ is directed along the
${\bf p}_\nu$ component which is perpendicular to ${\bf q}$
($\hat{{\bf e}}_\perp\cdot{\bf q}=0$). The sums in
Eqs.~(\ref{DCS_paral}) and~(\ref{DCS_perp}) run over all atomic
states $|n\rangle$, with $E_n$ being their energies, and, since
the ground state $|0\rangle$ is unpolarized, do not depend on the
direction of ${\bf q}$.

The representation~(\ref{DCS}) is similar to that for the case of
charged projectiles~\cite{fano63}. The longitudinal
term~(\ref{DCS_paral}) is associated with atomic excitations
induced by the force that the neutrino magnetic moment imposes on
electrons in the direction parallel to ${\bf q}$. And in the case
of nonrelativistic projectiles, only such excitations are of
significance. In the present case, however, the projectile is
ultrarelativistic. The transverse term~(\ref{DCS_perp})
corresponds to the exchange of a virtual photon which is polarized
as a real one, that is, perpendicular to ${\bf q}$. It can be
noted that it resembles a photoabsorption process when $Q\to T^2$
and the virtual-photon four-momentum thus approaches a physical
value, $q^2\to0$. An important feature is that, due to selections
rules, Eqs.~(\ref{DCS_paral}) and~(\ref{DCS_perp}) involve
different, nonintersecting sets of atomic states (see
Ref.~\cite{fano63} for detail). For this reason they do not
interfere. Another marked difference between the
terms~(\ref{DCS_paral}) and~(\ref{DCS_perp}) is that the first
vanishes in the forward direction, when $Q=T^2$, while the second
in the backward direction, when $Q=(2E_\nu-T)^2$.

In the reactor experiments one typically measures the DCS which
derives from~(\ref{DCS}) upon integrating over $Q$:
\begin{equation}
\label{DCS_dT}
\frac{d\sigma_{(\mu)}}{dT}=\int_{T^2}^{(2E_\nu-T)^2}\frac{d^2\sigma_{(\mu)}}{dTdQ}dQ
=\left(\frac{d\sigma_{(\mu)}}{dT}\right)_\parallel+\left(\frac{d\sigma_{(\mu)}}{dT}\right)_\perp.
\end{equation}
Below we separately analyze the longitudinal and transverse
components of Eq.~(\ref{DCS_dT}) assuming $E_\nu\gg T$, which
situation is usual for the reactor experiments.

\subsection{The longitudinal component of the differential cross section}
\label{longitude}
%
When integrating Eq.~(\ref{DCS_paral}) over $Q$, we neglect the
contribution of the term $\propto T^2/Q$ that brings only a minor
correction to the integral, so that~\cite{voloshin10}
\begin{equation}
\label{dsc_dt_paral}
\left(\frac{d\sigma_{(\mu)}}{dT}\right)_\parallel=\frac{\pi\alpha^2\mu_\nu^2}{m_e^2}
\int_{T^2}^{4E_\nu^2}S(T,Q)\frac{dQ}{Q},
\end{equation}
where we have introduced the so-called dynamical structure
factor~\cite{van_hove54}
\begin{equation}
\label{structure_factor}S(T,Q)=\sum_n|\langle n|\rho(-{\bf
q})|0\rangle|^2\delta(T-E_n+E_0).
\end{equation}
The latter is related to the density-density or polarization
Green's function $F$ as follows:
\begin{eqnarray}
\label{green} S(T,Q)&=&\frac{1}{\pi}{\rm Im}F(T+E_0,Q),
\nonumber\\
F(T+E_0,Q)&=&\langle0|\rho({\bf q})\frac{1}{T+E_0-H-i0}\rho(-{\bf
q})|0\rangle\nonumber\\
&=&\sum_n\frac{|\langle n|\rho(-{\bf
q})|0\rangle|^2}{T-E_n+E_0-i0},
\end{eqnarray}
where $H$ is the atomic Hamiltonian. Using
definitions~(\ref{structure_factor}) and~(\ref{green}), it is
straightforward to obtain the dispersion relation
\begin{equation}
\label{disp_rel}F(E,Q)=\frac{1}{\pi}\int_0^\infty\frac{{\rm
Im}F(E',Q)}{E-E'-i0}dE'.
\end{equation}
The dynamical structure factor~(\ref{structure_factor}) can be
expressed in terms of the generalized oscillator strengths
$f_n$~\cite{inokuti71},
\begin{eqnarray}
\label{osc_strength}S(T,Q)&=&\frac{Q}{2m_eT}\sum_n
f_n(Q)\delta(T-E_n+E_0), \nonumber\\
f_n(Q)&=&\frac{2m_e}{Q}(E_n-E_0)|\langle n|\rho(-{\bf
q})|0\rangle|^2,
\end{eqnarray}
and thus it satisfies the Bethe sum rule~\cite{inokuti71,bethe30}
\begin{equation}
\label{bethe_sum_rule}\int_0^\infty
S(T,Q)\frac{dT^2}{Q}=\frac{Z}{m_e}.
\end{equation}
However, it is not of much help for performing the integration
over $Q$ in Eq.~(\ref{dsc_dt_paral}). For the latter purpose the
following momentum-transfer dispersion relation was formulated in
Ref.~\cite{voloshin10} (cf. Eq.~(\ref{disp_rel})):
\begin{equation}
\label{disp_rel_voloshin}F(E,Q)=\frac{1}{\pi}\int_0^\infty\frac{{\rm
Im}F(E,Q')}{Q'-Q-i0}dQ',
\end{equation}
provided the energy $E$ is above the ionization threshold.
Consider the limit $Q\to0$. The electron density
operator~(\ref{density}) at ${\bf q}=0$ is by definition
$\rho(0)=Z$ and hence
$$
F(T+E_0,0)=\langle0|\rho(0)\frac{1}{T+E_0-H-i0}\rho(0)|0\rangle=\frac{Z^2}{T}.
$$
Using it in Eq.~(\ref{disp_rel_voloshin}) when $Q=0$, we arrive at
the sum rule (cf. Eq.~(\ref{bethe_sum_rule}))
\begin{equation}
\label{voloshin_sum_rule}\int_0^\infty
S(T,Q)\frac{dQ}{Q}=\frac{Z^2}{T}.
\end{equation}
Following the method of Ref.~\cite{voloshin10}, which implies the
use of Eq.~(\ref{voloshin_sum_rule}) for evaluating the integral
in Eq.~(\ref{dsc_dt_paral}) under assumptions of small $T$ and
large $E_\nu$, we get
\begin{equation}
\label{dcs_dT_voloshin}\left(\frac{d\sigma_{(\mu)}}{dT}\right)_\parallel\simeq\frac{\pi\alpha^2\mu_\nu^2}{m_e^2}\frac{Z^2}{T}
=Z^2\frac{d\sigma_{(\mu)}^{\rm FE}}{dT},
\end{equation}
where $d\sigma_{(\mu)}^{\rm FE}/dT$ is given by Eq.~(\ref{dcs_FE})
in the limit $T/E_\nu\to0$.

Thus, using the dispersion relation~(\ref{disp_rel_voloshin}), one
arrives at Eq.~(\ref{dcs_dT_voloshin}) where the factor of $Z^2$
occurs, which means that the atomic effects result in a coherent
enhancement of the DCS as compared to the case of $Z$ free
electrons, where a typical incoherent-scattering factor of $Z$ is
encountered (the same as, for instance, in the Compton
scattering). This indicates that Eq.~(\ref{disp_rel_voloshin}) is
not directly applicable to atoms with more than one electron. At
the same time, under some modification the method of
Ref.~\cite{voloshin10} can be viable, for instance, within the
mean-field model of many-electron atomic systems, where the
occupied one-electron orbitals independently contribute to the
$\nu II$ process. In such a case, one employs the sum
rule~(\ref{voloshin_sum_rule}) for each electron individually (the
proving of one-electron momentum-transfer sum rules is suggested
in Ref.~\cite{voloshin10}). Clearly, this procedure ensures a
physically-grounded factor of $Z$ in Eq.~(\ref{dcs_dT_voloshin})
after collecting together the one-electron contributions. Note
that the electron correlation effects beyond the mean-field
approximation crucially influencing multiple excitation and
ionization processes usually slightly affect those of single
ionization.


%
\subsection{The transverse component of the differential cross section}
\label{transverse}
As demonstrated below, it is the case which is relevant to the
work~\cite{wong10}. Integrating Eq.~(\ref{DCS_perp}) over $Q$ and
dropping the insignificant term $\propto Q/(2E_\nu-T)^2$, we get
\begin{equation}
\label{dsc_dt_perp}
\left(\frac{d\sigma_{(\mu)}}{dT}\right)_\perp=\frac{\pi\alpha^2\mu_\nu^2}{m_e^2}
\int_{T^2}^{4E_\nu^2}R(T,Q)\frac{dQ}{Q},
\end{equation}
where the function $R$ is defined as
\begin{equation}
\label{structure_current}R(T,Q)=\sum_n|\langle n|\hat{{\bf
e}}_\perp\cdot{\bf j}(-{\bf q})|0\rangle|^2\delta(T-E_n+E_0).
\end{equation}
Similarly to the dynamical structure
factor~(\ref{structure_factor}), it can be related to the
current-current Green's function $L$ (cf. Eq.~(\ref{green})):
\begin{eqnarray}
\label{green_current} R(T,Q)&=&\frac{1}{\pi}{\rm Im}L(T+E_0,Q),
\nonumber\\
L(T+E_0,Q)&=&\langle0|j_\perp({\bf
q})\frac{1}{T+E_0-H-i0}j_\perp(-{\bf q})|0\rangle\nonumber\\
&=&\sum_n\frac{|\langle n|j_\perp(-{\bf
q})|0\rangle|^2}{T-E_n+E_0-i0},
\end{eqnarray}
where $j_\perp(\pm{\bf q})=\hat{{\bf e}}_\perp\cdot{\bf j}(\pm{\bf
q})$.

Since ${\bf j}=(\rho{\bf v}+{\bf v}\rho)/2$, with ${\bf v}$ being
the electron velocity operator, we can roughly estimate the ratio
of the functions~(\ref{structure_current})
and~(\ref{structure_factor}) as $\sim\upsilon^2_a$, where
$\upsilon_a\ll1$ is a characteristic velocity of atomic electrons.
A more accurate estimate can be obtained when $Q$ is small on the
atomic scale. If $Q\ll2m_e\varepsilon_b$ one can treat
Eqs.~(\ref{structure_factor}) and~(\ref{structure_current}) in the
dipole approximation (see also Ref.~\cite{voloshin10}), which
yields
\begin{equation}
\label{dipole_approx}\frac{R(T,Q)}{S(T,Q)}=\frac{T^2}{Q}.
\end{equation}
Note that this ratio is much smaller than unity practically for
all $Q$ values involved in Eqs.~(\ref{dsc_dt_paral})
and~(\ref{dsc_dt_perp}). Thus, taking into account the foregoing
arguments, one might expect the transverse component to play a
minor role in Eq.~(\ref{DCS_dT}). The authors of
Ref.~\cite{wong10}, however, came to the contrary conclusion that
this component dramatically enhances due to atomic ionization when
$T\sim\varepsilon_b$. The enhancement mechanism proposed in
Ref.~\cite{wong10} is based on an analogy with the photoionization
process. As mentioned above, when $Q\to T^2$ the virtual-photon
momentum approaches the physical regime $q^2=0$. In this case, we
have for the integrand in Eq.~(\ref{dsc_dt_perp})
\begin{equation}
\label{photoeffect}\frac{R(T,Q)}{Q}{\Bigg|}_{Q\to
T^2}=\frac{\sigma_\gamma(T)}{4\pi^2\alpha T},
\end{equation}
where $\sigma_\gamma(T)$ is the photoionization cross section at
the photon energy $T$~\cite{akhiezer_book}. The limiting
form~(\ref{photoeffect}) was used in Ref.~\cite{wong10} in the
whole integration interval. Such a procedure is obviously
incorrect, for the integrand rapidly falls down as $Q$ ranges from
$T^2$ up to $4E_\nu^2$, especially when $Q\gtrsim r_a^{-2}$, where
$r_a$ is a characteristic atomic size (within the Thomas-Fermi
model $r_a^{-1}=Z^{1/3}\alpha m_e$~\cite{landafshiz_book}). This
fact reflects a strong departure from the real-photon regime. For
this reason we can classify the enhancement of the DCS determined
in Ref.~\cite{wong10} as spurious.

\section{Numerical estimates of electron binding effects}
\label{res}
\begin{figure}
\begin{center}
\includegraphics[width=0.80\textwidth]{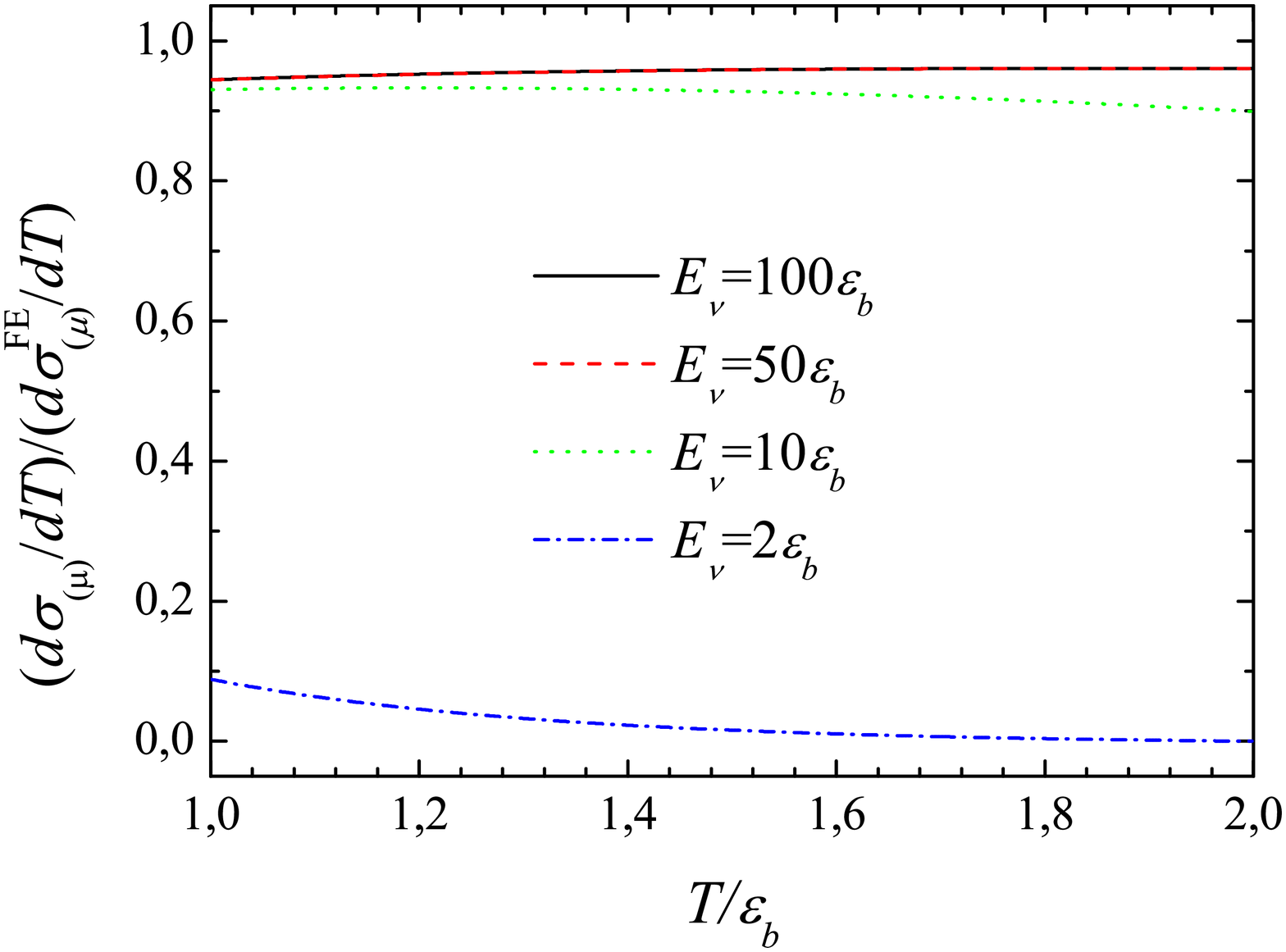}
\end{center}
\caption{\label{fig1} The DCS normalized to the FE result versus
$T/\varepsilon_b$ for different values of $E_\nu$. The
$E_\nu=50\varepsilon_b$ and $E_\nu=100\varepsilon_b$ curves are
practically indistinguishable.}
\end{figure}

Let us inspect numerically the magnetic $\nu II$ process of a
hydrogenlike ion with a nuclear charge $Z=32$ that mimics the case
of a $K$-electron in Ge, which is a detector material in the
reactor experiments~\cite{texono07,gemma10}. The
functions~(\ref{structure_factor}) and~(\ref{structure_current})
are calculated analytically using the well-known expressions for
the generalized oscillator strengths~(\ref{osc_strength}) (see,
for instance, Ref.~\cite{inokuti71}) and photoionization matrix
elements~\cite{akhiezer_book} corresponding to transitions from a
$1s$ hydrogenlike state. Then, the $Q$ integrations in
Eq.~(\ref{DCS_dT}) are performed numerically. Fig.~\ref{fig1}
shows the DCS~(\ref{DCS_dT}), which is normalized to the FE
value~(\ref{dcs_FE}), above the ionization threshold as a function
of $T/\varepsilon_b$, with the electron binding energy given by
$\varepsilon_b=\alpha^2Z^2m_e/2\approx14$~keV. As can be seen, the
AE results for $E_\nu\gg\varepsilon_b$ are close to the FE ones in
magnitude.
Such a behavior agrees both with the analytical estimates
performed in Ref.~\cite{voloshin10} and with the numerical results
of more rigorous, relativistic treatments for various atomic
targets~\cite{kurchatnik}. It can be qualitatively explained by
noticing the following facts. First, in an attractive Coulomb
potential there is an infinite set of bound states, with the
discrete spectrum smoothly transforming into the continuum at the
ionization threshold. Second, the average value of the
$K$-electron momentum is $p_e=\alpha Zm_e$ and the average change
in the electron momentum after ejection, $\Delta p_e$, is such
that $\Delta p_e^2=2m_eT$, which is analogous to the FE case.
Thus, in contrast to the claim of Ref.~\cite{wong10}, at small
recoil-electron energies one might expect no enhancement of the
sensitivity of the experiments searching for $\mu_\nu$.

\begin{figure}
\begin{center}
\includegraphics[width=0.80\textwidth]{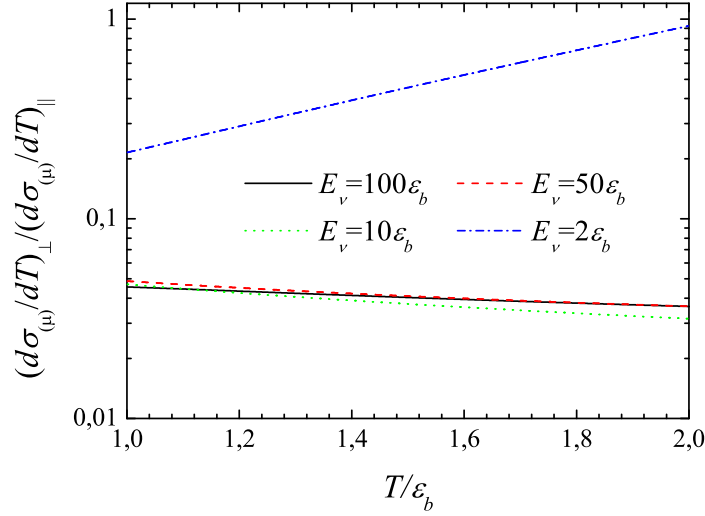}
\end{center}
\caption{\label{fig2} The dependence of the ratio between the
transverse and longitudinal components of the DCS~(\ref{DCS_dT})
on $T/\varepsilon_b$ at different values of $E_\nu$.}
\end{figure}

Fig.~\ref{fig2} presents numerical results for the ratio of the
transverse and longitudinal terms of Eq.~(\ref{DCS_dT}) in the
same kinematical situations as in Fig.~\ref{fig1}. As anticipated,
when $E_\nu\gg\varepsilon_b$ the relative role of the transverse
component is subsidiary. Though the results for
$E_\nu=2\varepsilon_b$ are not strictly relevant to the reactor
experiments, where the typical neutrino energies are much higher
($\sim1$~MeV), they are shown for illustrative purposes--in this
case the contribution from the transverse excitations becomes
considerable, especially as $T$ increases. This finding can be
explained by Eq.~(\ref{dipole_approx}) and different kinematical
factors $1-T^2/Q$ and $1-Q/(2E_\nu-T)^2$ in Eqs.~(\ref{DCS_paral})
and~(\ref{DCS_perp}), respectively. It also can be noted that the
kinematical regime where $T$ is close to $E_\nu$ validates
applicability of the real-photon approximation~(\ref{photoeffect})
in the whole kinematically available region $T^2\leq
Q\leq(2E_\nu-T)^2$. However, as clearly seen from Fig.~\ref{fig1},
even in a such marginal situation the DCS behaves oppositely to
the resonant-enhancement scenario of Ref.~\cite{wong10}.

\section{Conclusions}
\label{concl}
To summarize, we have performed a theoretical analysis of the
magnetic neutrino scattering on atomic electrons. For this purpose
we have divided the DCS into two components corresponding to the
longitudinal and transverse atomic excitations. Incorrectness of
the recent theoretical predictions assigning the atomic effects to
play a significant role in the magnetic neutrino
scattering~\cite{wong10} has been revealed. Numerical calculations
of the DCS for the magnetic $\nu II$ process of a hydrogenlike
Ge$^{+31}$ ion have been carried out at different neutrino
energies. The results exhibited suppression of the DCS relative to
the FE values, which is slight at high impact energy and
pronounced when the latter is comparable to the electron binding
energy.

The present results based on nonrelativistic calculations
qualitatively agree with those using a relativistic
Dirac-Hartree-Fock treatment of atomic
electrons~\cite{kurchatnik}. No enhancement of the DCS due to
electron binding effects has been determined, in contrast to
Ref.~\cite{wong10}. In this regard, the analytical approach
arguing the insignificance of the atomic effects~\cite{voloshin10}
might have a rich potential, provided that it is appropriately
modified. Finally, it is unreasonable to expect the effects of
atomic excitation and/or ionization to introduce any enhancement
of the sensitivities of the experiments searching for neutrino
magnetic moments. Therefore, it will be interesting to explore the
role of coherent magnetic neutrino scattering on atoms in
detectors, which case, however, requires much lower energy
thresholds in the deposited energy $T$ ($\sim100$~eV) than
presently attainable in the detectors ($\sim1$~keV).

\section*{Acknowledgements}
We thank Victor B. Brudanin and Alexander S. Starostin for useful
discussions. We are grateful to Mikhail B. Voloshin for valuable
comments.
%






\begin{thebibliography}{00}
%
\bibitem{giunti09}C. Giunti and A.I. Studenikin, Phys. At. Nucl.
73 (2009) 2089, arXiv:08123646 [hep-ph].
%
\bibitem{marciano77}W.J. Marciano and A.I. Sanda, Phys. Lett. B 67 (1977)
303.
%
\bibitem{lee77}B.W. Lee, R.E. Shrock, Phys. Rev. D 16 (1977)
1444.
%
\bibitem{fujikawa80}K. Fujikawa, R.E. Shrock, Phys. Rev. Lett. 45 (1980) 963.
%
\bibitem{munu05}Z. Daraktchieva et al. (MUNU collaboration), Phys.
Lett. B 615 (2005) 153, arXiv:hep-ex/0502037.
%
\bibitem{texono07}H.T. Wong et al. (TEXONO collaboration), Phys. Rev. D 75 (2007) 012001.
%
\bibitem{gemma10}A.G. Beda et al. (GEMMA collaboration), arXiv:0906.1926 [hep-ex]; arXiv:1005.2736 [hep-ex].
%
\bibitem{raffelt90}G.G. Raffelt, Phys. Rev. Lett. 64 (1990) 2856.
%
\bibitem{domogatskii71}G.V. Domogatskii, D.K. Nadezhin, Sov. J.
Nucl. Phys. 12 (1971) 678.
%
\bibitem{vogel89}P. Vogel, J. Engel, Phys. Rev. D 39 (1989) 3378.
%
\bibitem{wong10}H.T. Wong, H.-B. Li, S.-T. Lin, Phys. Rev. Lett. 105 (2010) 0161801, arXiv:1001.2074 [hep-ph].
%
\bibitem{voloshin10} M.B. Voloshin, Phys. Rev. Lett. 105 (2010) 201801, arXiv:1008.2171 [hep-ph].
%
\bibitem{fano63} U. Fano, Annu. Rev. Nucl. Sci. 13 (1963) 1.
%
\bibitem{van_hove54}L. Van Hove, Phys. Rev. 95 (1954) 249.
%
\bibitem{inokuti71}M. Inokuti, Rev. Mod. Phys. 43 (1971) 297.
%
\bibitem{bethe30}H. Bethe, Ann. Physik 5 (1930) 325.
%
\bibitem{akhiezer_book}A.I. Akhiezer, V.B. Berestetskii, Quantum Electrodynamics, second
ed., Wiley, New York, 1965.
%
\bibitem{landafshiz_book}L.D. Landau, E.M. Lifshitz, Quantum Mechanics, Non-Relativistic Theory, third
ed., Pergamon, New York, 1977.
%
\bibitem{kurchatnik}V.Yu. Dobretsov, A.B. Dobrotsvetov, S.A. Fayans, Sov. J. Nucl. Phys. 55
(1992) 1180; V.I. Kopeikin, L.A. Mikaelyan, V.V. Sinev, S.A.
Fayans, Phys. At. Nucl. 60 (1997) 1859; S.A. Fayans, L.A.
Mikaelyan, V.V. Sinev, Phys. At. Nucl. 64 (2001) 1475.
%


%
\end{thebibliography}



%

%

\end{document}